\documentclass{elsart}
\usepackage{graphicx}
\usepackage{amsmath,amssymb}

\begin{document}
\begin{frontmatter}
\title{Some exact results for the trapping of subdiffusive particles
in one dimension}
\author{S. B. Yuste\corauthref{sby}}
\corauth[sby]{Corresponding author.} \ead{santos@unex.es}
\ead[url]{http://www.unex.es/fisteor/santos/sby}
\author{ L. Acedo}

\address{Departamento de F\'{\i}sica, Universidad  de  Extremadura,
E-06071 Badajoz, Spain}
\begin{abstract}
We study a generalization of the standard trapping problem of
random walk theory in which particles move \emph{subdiffusively}
on a one-dimensional lattice. We consider the cases in which the
lattice is filled with a one-sided and a two-sided random
distribution of static absorbing traps with concentration $c$. The
survival probability $\Phi(t)$ that the random walker is not
trapped by time $t$ is obtained exactly in both versions of the
problem through a fractional diffusion approach. Comparison with
simulation results is made.
\end{abstract}
\end{frontmatter}
\maketitle
\section{Introduction}
\label{sect_1} The trapping of Brownian particles by static traps
randomly distributed over either a Euclidean or a disordered
substrate is a fundamental problem of non-equilibrium statistical
mechanics and chemistry with a very wide range of applications
\cite{Hughes,Weiss,HollanderWeiss,ShDba,AvrahamHavlinDifuReacBook}.
This is also one of the oldest problems in random walk theory
which was essentially formulated by Smoluchowski at the beginning
of past century in his theory of coagulation of colloidal
particles
\cite{Hughes,Weiss,HollanderWeiss,AvrahamHavlinDifuReacBook}. This
model has proven useful in research areas such as the trapping of
mobile defects in crystals with point sinks
\cite{Beeler,Rosens,Damask}, the kinetics of luminescent organic
materials \cite{Rosens}, the kinetics of photosynthetic light
energy to oxygen conversion \cite{Montroll}, anchoring of polymers
by chemically active sites \cite{Oshanin}, atomic diffusion in
glasslike materials \cite{Miyagawa}, and others \cite{Klafter}.
Some generalizations of the basic model have also been considered
recently. The case of a gated (ungated) random walker trapped by a
distribution of ungated (gated) fixed traps has been proposed in
connection with the kinetics of reactions between complex
molecules whose active groups are screened as the molecules
diffuse \cite{PREBenichou}. The trapping problem with many random
walkers has also been studied in one dimension \cite{Multipart}
and the corresponding trapping statistics have been found on
Euclidean and fractal lattices
\cite{TRNReview,OrderEucl,OrderFrac}. The multiparticle
predator-prey problems in which a single fixed trap, the ``lamb'',
is captured by one of a set of $N$ random walkers or ``lions''
initially placed at a given distance from the prey was also
discussed by  Krapivsky and Redner \cite{KR,KR-Am}.  Another quite
interesting and difficult variation of the standard trapping
problem is that where the traps diffuse too
\cite{AvrahamHavlinDifuReacBook,movinTraps}.

 Trapping reactions between molecules embedded in
biological samples and disordered materials are usually
handicapped by the porous and statistical fractal structure of
these media \cite{ShDba}. In some cases this gives rise to
subdiffusion of the particles, i.e., the mean square displacement
$ \langle r^2 (t) \rangle$ of the particles from the original
starting site is no longer linear on time, but verifies a
generalized Fick's second law:
\begin{equation}
\label{r2t} \langle r^2 (t) \rangle \approx \displaystyle\frac{2
K_\gamma}{\Gamma(1+\gamma)} t^\gamma
\end{equation}
where $\gamma$ (with $0 < \gamma < 1$) is the (anomalous)
diffusion exponent and $K_\gamma$ is the diffusion coefficient. Of
course, there are many other instances in which subdiffusion
processes appear
\cite{BouchaudPhysRep90,SubdifuRandPot,KantorCM,Klemm,Kosztolowicz}.
A useful approach for understanding  subdiffusion processes is by
means of the continuous time random walk model (CTRW) in which the
random walker performs jumps with a waiting time distribution with
a broad long-time tail: $\psi(t) \sim t^{-(1+\gamma)}$ for large
$t$
\cite{Hughes,Weiss,HollanderWeiss,BouchaudPhysRep90,Rangarajan,MetzRev}.
The long-time tail of this waiting time distribution incorporates,
in a statistical sense, the effect of the bottlenecks and the
dead-ends in the diffusion of the random walker embedded in the
disordered structure, and the model is compatible with Eq.\
(\ref{r2t}). For subdiffusive random walkers the continuum
description given by the ordinary diffusion equation is replaced
by the fractional diffusion equation \cite{MetzRev}
\begin{equation}
\label{subdeq} \frac{\partial}{\partial t} W(x,t)=K_\gamma \;{}_0
D_t^{1-\gamma}\, \frac{\partial^2}{\partial x^2} W(x,t)\; ,
\end{equation}
where ${}_0 D_t^{1-\gamma}$ is the Riemann-Liouville fractional
derivative of order $1-\gamma$ \cite{MetzRev,Podlubny,Hilfer},
\begin{equation}
~_{0}\,D_{t}^{1-\gamma } W(x,t)=\frac{1}{\Gamma(\gamma)}
\frac{\partial}{\partial t} \int_0^t d\tau
\frac{W(x,\tau)}{(t-\tau)^{1-\gamma}},
\end{equation}
 and $W(x,t)$ is the probability density that the particle that started at $0$ at time
$0$ is at $x$ at time $t$.

Our objective in this paper is to find analytical expressions for
the survival probability $\Phi(t)$  defined as the probability
that no trap site has been reached by the \emph{subdiffusive}
random walker by time $t$. Two variations of the classical
trapping problem (sometimes called Rosenstock's trapping problem)
are considered: (i) the ``one-sided''  trapping problem
\cite{Multipart} in which only  {\em one half-line} of a
one-dimensional lattice is filled with a random distribution of
static traps (this process could mimic the excitation or
production of defects on one side of a fiber by irradiation, the
other side being shielded); and (ii) the ``two-sided'' trapping
problem corresponding to the trapping of a single subdiffusive
random walker placed initially at $x=0$ between  {\em two
half-lines} ($x < 0$ and $x > 0$) randomly filled with static
traps. We find exact analytical solutions for both problems and
compare them with simulations. As precedents of these results we
whould cite the work of Blumen, Klafter and Zumofen (see
\cite{Klafter,Zumofen,BKZRev} and references therein) about the
trapping of particles on Euclidean and fractal substrates for
random walkers with waiting time distributions with broad
long-time tails: $\psi(t) \sim t^{-1-\gamma}$, $0<\gamma<1$.

The paper is organized as follows. In Sec.~\ref{sec:onesided}
 the exact solution of the one-sided  trapping
problem is obtained and compared with simulation results. In
 Sec.~\ref{sec:Rosenstock} we build the extended Rosenstock's
approximation for the survival probability of the particle in the
one-sided trapping problem  by calculating an exact expression for
the moments of the one-sided span (the number of distinct sites
visited by the random walker in a given direction). In
Sec.~\ref{sec:twosided} we obtain the exact survival probability
of the subdiffusive random walker for the two-sided case in
integral form. This integral is evaluated in the asymptotic
long-time and short-time limits in order to get simple closed
expressions. The paper ends with some conclusions and remarks in
 Sec.~\ref{sec:conclusions}.

\section{The one-sided trapping problem}
\label{sec:onesided} 
\begin{figure}
\includegraphics[width=0.95 \columnwidth]{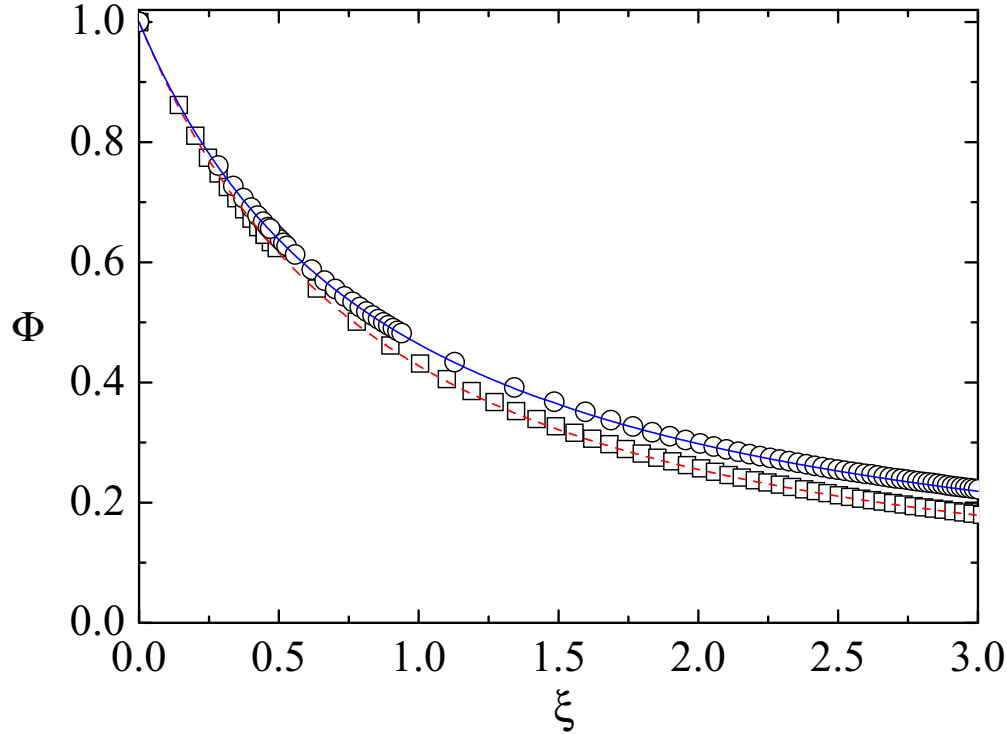}
\caption{Survival probability $\Phi(t)$ versus $\xi=\lambda
\sqrt{K_\gamma t^\gamma}$. The circles are simulation results for
subdiffusive walkers with $\gamma=1/2$. The squares are simulation
results for normal diffusive walkers with $\psi(t)=e^{-t}$. The
lines are the corresponding exact results given by Eqs.
\eqref{PhitSub} (solid line) and \eqref{PhiDif} (dashed line),
respectively.
 \label{fig1}}
\end{figure}

In the one-sided trapping model quenched
traps are randomly distributed on, say, the right-hand side of a
one-dimensional lattice ($x > 0$) with concentration $c$. The
random walker is placed initially upon site $x=0$ and  performs
jumps to its  nearest neighbour sites with a waiting time
distribution $\psi(t)$ until it reaches a trap site where it is
absorbed. The survival probability is given by \cite{Hughes}:
\begin{equation}
\label{PhiSum} \Phi(t)=\sum_{r=1}^t  e^{-\lambda r} P(t \vert r)
\; ,
\end{equation}
where $\lambda=-\ln(1-c)$ and  $P(t \vert r)$ is the probability
that the span of the random walker in the positive direction (the
largest distance reached by the random walker for $x > 0$) is
equal to $r$ after $t$ time steps. Now let $\Gamma(t \vert r)$ be
the probability that the site $x=r$ has not been visited by the
random walker by time $t$ (the so-called fixed-trap survival
probability). Because \cite{SA}
\begin{equation}
\label{spanden}
P(t \vert r)=\displaystyle\frac{d \Gamma(t \vert
r)}{d r}\; ,
\end{equation}
Eq.\ (\ref{PhiSum}) takes the following form in the continuous
limit:
\begin{equation}
\label{PhiInt} \Phi(t)=\int_0^\infty \, e^{-\lambda r} \frac{d
\Gamma(t \vert r)}{d r} d r\; ,
\end{equation}
or, integrating by parts,
\begin{equation}
\label{PhiInt1} \Phi(t)=\lambda \int_0^\infty e^{-\lambda r}
\Gamma(t \vert r) d r\;,
\end{equation}
 because $\Gamma(t \vert 0)=0$, $\Gamma(t \vert \infty)=1$ and
$e^{-\lambda r} \rightarrow 0$ as $r \rightarrow \infty$. For
subdiffusive particles, the function $\Gamma(t|r)$ can be written
in terms of Fox's $H$ function  \cite{Metzler}:
\begin{equation}
\label{GFixTrap}
\Gamma(t \vert r)=1-
H^{10}_{11}\left[\frac{r}{\sqrt{K_\gamma} t^{\gamma/2}}
\left|\begin{array}{l}{(1 ,\gamma/2)}\\[1ex]{(0,1)}
\end{array}\right.\right] \; .
\end{equation}
The time Laplace transform of $\Gamma(t \vert r)$ is given by
\begin{equation}
\label{LapGFix} \widetilde{\Gamma}(s \vert
r)=\frac{1}{s}\left[1-\text{exp}(-r \sqrt{s^\gamma})\right]\; .
\end{equation}
Then the Laplace transform of $\Phi(t)$ is readily calculated from
Eqs.\ (\ref{PhiInt1}) and (\ref{LapGFix}):
\begin{equation}
\label{PhiLapl1} \widetilde{\Phi}(s)=
\frac{1}{s+\lambda\sqrt{K_\gamma } s^{1-\gamma/2}} \;
\end{equation}
so that \cite{Erdely}
\begin{equation}
\label{PhitSub} \Phi(t)= E_{\gamma/2}(-\xi)\; ,
\end{equation}
where $\xi \equiv \lambda \sqrt{K_\gamma t^\gamma}$ and
$E_\alpha(z)$ is the Mittag-Leffler function with parameter
$\alpha$ \cite{Erdely,Podlubny}. For $\gamma=1$, the
Mitagg-Leffler function becomes \cite{MetzRev,Podlubny}:
\begin{equation}
\label{PhiDif} \Phi(t)=e^{\xi^2} \text{erfc} \left( \xi \right)\;
,
\end{equation}
with $\xi=\lambda \sqrt{D t}$, ($K_1\equiv D$), and we recover the
result for the survival probability of a normal diffusive random
walker in the one-sided trapping problem \cite{Multipart}. For
very long times, the asymptotic expansion of the Mittag-Leffler
function \cite{Podlubny} allows us to write
\begin{equation}
\label{PhiN1as} \Phi(t)=\sum_{k=1}^n
\frac{(-1)^{k+1}}{\Gamma(1-k\gamma/2) } \xi^{-k}
+\mathcal{O}(\xi^{-1-n})\; .
\end{equation}
Thus, an asymptotic time regime is reached for $t \gg 1/(K_\gamma
\lambda^2)^{1/\gamma}$ where the survival probability exhibits a
power law decay
\begin{equation}
\label{OSDVlim} \Phi(t) \approx \frac{1}{\Gamma(1-\gamma/2)}\frac{
1}{\lambda  \sqrt{\,K_\gamma \, t^{\gamma}}}\; .
\end{equation}
This is an {\em algebraic} fluctuation slowdown corresponding to
the Donsker-Varadhan limit \cite{Hughes,Weiss,DV}. We compare in
Fig.~\ref{fig1} the exact survival probability for a subdiffusive
random walker with a waiting time distribution with the form of
the Pareto law
\begin{equation}
\label{psit}
\psi(t)=\displaystyle\frac{\gamma}{(1+t)^{1+\gamma}}\;
\end{equation}
and  $\gamma=1/2$. Taking the distance between nearest neighbour
sites (jump length) as 1, the subdiffusion constant is
$K_{1/2}=1/\left[2\Gamma(1/2)\right]=1/\sqrt{4 \pi}$. The
concentration of traps used was $c=0.01$.

\section{The Rosenstock's approximation for the one-sided trapping problem}
\label{sec:Rosenstock}

Let  $S(t)$ be the number of distinct sites on the positive
half-line visited up to time $t$ by a random walker who started at
$x=0$ at time $t=0$. Then, the  survival probability of the random
walker is given by  $\Phi(t)=\left\langle (1-c)^{S(t)}
\right\rangle = \left\langle e^{- \lambda S(t)} \right\rangle$,
the average being performed over all realizations of the random
walker's exploration of the lattice from time $0$ until time $t$.
By means of the cumulant expansion technique \cite{Hughes,Weiss}
the following equivalent form is derived:
\begin{equation}
\label{Ros:gen} \Phi(t)=\exp \left[\sum_{n=1}^\infty
\displaystyle(-1)^n \frac{\kappa_n \lambda^n}{n !}\right] \; ,
\end{equation}
where $\kappa_n$, $n=1,2,\ldots$ denote the cumulants of $S(t)$:
$\kappa_1 = \left\langle S(t) \right\rangle$,
$\kappa_2=\left\langle S^2(t) \right\rangle -\left\langle S(t)
\right\rangle^2 \equiv \sigma^2(t)$, $\ldots$. If we keep the
first $n+1$ terms of the sum in equation (\ref{Ros:gen}) we arrive
at the $n$th-order Rosenstock approximation
\cite{Hughes,Weiss,Zumofen}. The error made by using this
approximation is $\mathcal{O}(\kappa_{n+2} \lambda^{n+2})$. Thus,
the condition $\kappa_{n+2} \ll 1/\lambda^{n+2}$ must be fulfilled
for the $n$th-order Rosenstock approximation to be reasonable.

For the Rosenstock approximation to be useful for the subdiffusive
case, it is necessary to know the cumulants or, equivalently, the
moments of $S(t)$ for subdiffusive particles. The next step is
thus to derive exact expressions for the moments $\langle
S^m(t)\rangle$, $m=1,2,\ldots$ of the one-sided span  of a
subdiffusive random walker, i.e., the territory explored in a
given direction. The probability density of the one-sided span,
$P(t\vert r)$, was given in Eq.\ (\ref{spanden}) in terms of the
survival probability. Hence
\begin{equation}
\label{momenS+1} \langle S^m(t) \rangle = \int_0^\infty
\frac{d\Gamma(t|r)}{dr}r^m dr= -\int_0^\infty \frac{d}{d r}
(1-\Gamma(t|r)) r^m dr \;
\end{equation}
or, integrating by parts,
\begin{equation}
\label{momenS1} \langle S^m(t)\rangle = m \int_0^\infty
\left[1-\Gamma(t|r)\right]r^{m-1} dr \; ,
\end{equation}
where we have taken into account that $\Gamma(t | 0)=0$ and
$\lim_{r\rightarrow \infty} r^m [1-\Gamma(t | r)]=0$ as a
consequence of the stretched exponential behaviour of
$1-\Gamma(t|r)$ as $r \rightarrow \infty$ for subdiffusive
particles \cite{Metzler}. From Eq.\ (\ref{momenS1}) and the
explicit expression of the survival probability in Eq.\
(\ref{GFixTrap}) we obtain
\begin{equation}
\label{S1momen} \left\langle S^m(t) \right\rangle=m
s_m\left(K_\gamma t^\gamma \right)^{m/2}\; ,
\end{equation}
with
\begin{equation}
\label{sm}
 s_m=\int_0^\infty
z^{m-1}H^{10}_{11}\left[z
\left|\begin{array}{l}{(1 ,\frac{\gamma}{2})}\\[1ex]{(0,1)}\end{array}\right.\right] dz\; .
\end{equation}
In order to calculate the coefficients $s_m$, $m=1,2,\ldots$ we
observe that the Laplace transform of the Fox $H$ function
appearing in the integrand in Eq.\ (\ref{sm}), $\widetilde H(u)$,
is a generating function of these coefficients:
\begin{equation}
\label{gen}
 \widetilde H(u) =\sum_{m=1}^\infty (-1)^{m-1}
\frac{u^{m-1}}{(m-1)!}\; s_m\; .
\end{equation}
Taking into account some properties of the Fox functions
\cite{West,Mathai,MetzRev} we finally identify $\widetilde H(u)$
with a two-parameter Mittag-Leffler function as follows:
\begin{equation}
\label{Hu}
 \widetilde H(u)=E_{\gamma/2,1+\gamma/2}(-u)\; ,
\end{equation}
where, in the last identity, we have used the relation
\cite{MetzRev}
\begin{equation}
\label{EabeqH} E_{\alpha\beta}(-u)=H^{11}_{12}\left[u
\left|\begin{array}{l}{(0
,1)}\\[1ex]{(0,1),(1-\beta,\alpha)}\end{array}\right.\right]\;
\end{equation}
between the two-parameter Mittag-Leffler functions and the Fox
functions \cite{MetzRev}. The function $E_{\alpha\beta}(u)$ admits
the following series expansion \cite{MetzRev,Podlubny}:
\begin{equation}
\label{Eabser} E_{\alpha\beta}(u)=\displaystyle\sum_{k=0}^\infty
\frac{z^k}{\Gamma(\alpha k+\beta)}\, , \qquad \alpha>0,\quad
\beta>0\; .
\end{equation}
From Eqs.\ \eqref{gen}, (\ref{Hu}) and (\ref{Eabser}) we find
\begin{equation}
\label{smresult} s_m =\frac{(m-1)!}{\Gamma(1+m\gamma/2)}\, , \quad
m=1,2,\ldots\;
\end{equation}
so that from Eq.\ (\ref{S1momen}) we finally obtain
\begin{equation}
\label{SExploram} \left\langle S^m(t) \right\rangle
=\frac{m!}{\Gamma(1+m\gamma/2)} (K_\gamma t^\gamma)^{m/2}\; .
\end{equation}
Obviously, for $\gamma=1$ we recover the exact result for the
moments of the one-sided span of a diffusive random walker
\cite{Multipart}:
\begin{equation}
\left\langle S^m(t) \right\rangle
=\frac{\Gamma[(m+1)/2]}{\sqrt{\pi}} (4Dt)^{m/2}\; .
\end{equation}
From Eq.\ (\ref{SExploram}) one can calculate the cumulants
$\kappa_n$, $n=1$,$2$,$\ldots$ of the one-sided span distribution,
and their direct substitution into Eq.\ \eqref{Ros:gen} yields the
general expression for the Rosenstock approximation for
subdiffusive particles:
\begin{equation}
\label{Phi1:Ros} \Phi(t)=\exp\left\{ \sum_{n=1}^\infty
\frac{(-1)^n}{n!}  a_n \xi^n \right\}\; .
\end{equation}
The first three  coefficients $a_n$ are:
\begin{equation}
\begin{array}{rcl}
a_1&=&\displaystyle\frac{1}{\Gamma(1 + \frac{\gamma }{2})}\\
\noalign{\smallskip} a_2&=&-\displaystyle\frac{1}{[{{\Gamma}(1 +
\frac{\gamma }{2})}]^{2} } +
  \frac{2}{{\Gamma}(1 + \gamma )}\\
\noalign{\smallskip} a_3&=& \displaystyle\frac{2}{[{\Gamma}(1 +
\frac{\gamma }{2})]^{3}} -
    \frac{6}{{\Gamma}(1 + \frac{\gamma }{2})\,
       {\Gamma}(1 + \gamma )} +
    \frac{6}{{\Gamma}(1 + \frac{3\,\gamma }{2})}\; .
\end{array}
\end{equation}
It is interesting to note that from Eqs.\ (\ref{PhitSub}) and
(\ref{Phi1:Ros}) we get the following identity
\begin{equation}
\label{lnEg} \ln E_{\gamma/2}(\xi)= \sum_{n=1}^\infty \frac{
a_n}{n!} \; \xi^n\; ,
\end{equation}
which provides a series expansion of the logarithm of the
Mittag-Leffler function.

\section{The two-sided trapping problem}
\label{sec:twosided}

In this section we calculate the exact survival probability of a
single subdiffusive random walker placed initially  between two
half-lines populated with a random distribution of traps with
concentration $c$. We will use parallel arguments to the standard
ones for diffusive random walkers \cite{Weiss,Redner}. We must
also mention that an alternative approach was proposed by Anlauf
\cite{Anlauf} to find the long-time behavior for the
one-dimensional trapping problem with a diffusive particle.

As a starting point we evaluate  the probability $W(x,t \vert
x_0,t=0)$ that a random walker starting from $x_0$ at $t=0$ inside
a box $0 \le x \le L$ with absorbing boundaries is at $x$ at time
$t$. The function $W(x,t\vert x_0,t=0)$ satisfies the fractional
partial differential equation (the subdiffusion equation):
\begin{equation}
 \frac{\partial }{\partial t} W(x,t|x_0,0)= K_\gamma
~_{0}\,D_{t}^{1-\gamma } \frac{\partial^2}{\partial x^2}
W(x,t|x_0,0)\; , \label{Pfracdifu}
\end{equation}
with the boundary conditions $W(0,t\vert x_0, 0)=W(L,t\vert x_0,
0)=0$, and the initial condition $W(x,0 \vert x_0,
0)=\delta(x-x_0)$ with $0<\gamma\leq 1$. Equation
(\ref{Pfracdifu}) is straightforwardly solved by separation of
variables \cite{MetzRev}:
\begin{equation}
\label{solx1mx} W(x,t|x_0,0)=\frac{2}{L} \sum_{n=1}^\infty
\sin\left(\frac{n\pi x_0}{L}\right) \sin\left(\frac{n\pi
x}{L}\right) E_\gamma\left(-K_\gamma \frac{n^2\pi^2}{L^2}
t^\gamma\right)\; ,
\end{equation}
where we have taken into account that the solution of the
fractional differential equation $ d T/d t=-K_\gamma \lambda^2
~_{0}\,D_{t}^{1-\gamma } T$ is given by a Mittag-Leffler function
$T(t)=E_\gamma\left(-K_\gamma \lambda^2 t^\gamma \right)$. The
survival probability of a subdiffusive random walker starting from
$x_0 \in (0,L)$ is given by $\int_0^L W(x,t \vert x_0,0)dx$. Let
$\Phi_L(t)$ be the survival probability of the random walker
averaged over all configurations of traps which contain the origin
inside a hole of length $L$. Then $\Phi_L(t)=\int_0^L \int_0^L
W(x,t|x_0,0) \,dx \,dx_0$, and Eq.~\eqref{solx1mx} yields
\begin{equation}
\label{PhiL} \Phi_L(t)=\displaystyle\frac{8}{\pi^2}
\displaystyle\sum_{n=0}^\infty \displaystyle\frac{1}{(2n+1)^2}\,
 E_\gamma\left(-\frac{(2n+1)^2\pi^2}{L^2}
\lambda^2 K_\gamma t^\gamma\right)\; .
\end{equation}
 The survival probability of the subdiffusive random
walker in the two-sided trapping problem is finally given by an
average of $\Phi_L(t)$ over the distribution of hole lengths in
the random trap configurations. This distribution is
$\eta(L)=\lambda^2 L e^{-\lambda L}$ \cite{Weiss}, and
consequently we have
\begin{align}
\Phi(t)&= \lambda^2 \int_0^\infty L
e^{-\lambda L} \Phi_L(t)dL \nonumber\\
&=\frac{8\lambda^2}{\pi^2} \sum_{n=0}^\infty
\frac{1}{(2n+1)^2}\,\int_0^\infty L\, e^{-\lambda L}
E_\gamma\left(- \frac{\beta_n}{L^2} \right) dL\;
 \label{Phitavg}
\end{align}
with $\beta_n=K_\gamma (2n+1)^2\pi^2 t^\gamma$, $n=1$, $2$,
$\ldots$. This is the exact integral representation of the
survival probability of the subdiffusive particle in the two-sided
trapping problem.

\subsection{Long-time behavior of $\Phi(t)$}

 For $\gamma=1$ the Mittag-Leffler function reduces to an
exponential and the integrals in Eq.\ \eqref{Phitavg} are
estimated asymptotically for large times by means of the Laplace
method \cite{Weiss}. However, this approach is not possible for
$\gamma \neq 1$. In order to perform a long-time asymptotic
evaluation of the integrals in Eq.\ (\ref{Phitavg}), we use the
following series expansion of the Mittag-Leffler function:
\begin{equation}
\label{MLserie}
 E_\gamma\left(-
\displaystyle\frac{\beta_n}{L^2} \right)=\displaystyle\sum_{m=1}^p \displaystyle\frac{(-1)^{m+1}
L^{2m}}{\beta_n^m \Gamma(1-\gamma
  m)}+O\left(\frac{L^2}{\beta_n}\right)^{1+p}\; ,
\end{equation}
which is valid for $\beta_n/L^2\rightarrow\infty$ in the range $0
< \gamma < 2$ \cite{Podlubny}. Inserting this series expansion
into Eq.\ (\ref{Phitavg}), integrating term by term, and after
some rearrangements we find
\begin{equation}
\label{Philarget} \Phi(t)=\displaystyle\frac{2}{\pi^2}
\displaystyle\sum_{m=1}^p  \displaystyle\frac{
\left(4-2^{-2m}\right)\zeta(2m+2) \Gamma(2m+2)}{(-1)^{m+1}
\Gamma(1-\gamma
  m) \pi^{2m}\xi^{2m}} +
  O\left( \xi^{-2p-2}\right)\; ,
\end{equation}
where we have taken into account that $\sum_{n=0}^\infty\,
1/(2n+1)^{2m+2}=(1-2^{-2m-2})\zeta(2m+2)$, $\zeta(s)$ being
Riemann's zeta function. The dominant term is
\begin{equation}
\label{longtime}
\Phi(t) \sim \displaystyle\frac{1}{2 \Gamma(1-\gamma) \lambda^2 K_\gamma}\, t^{-\gamma}\, ,
\end{equation}
which is valid for $\xi \gg 1$. Therefore we find an algebraic
fluctuation slowdown of the survival probability corresponding to
the Donsker-Varadhan \cite{DV} limit with an exponent two times
greater than that of the one-sided long-time behavior in Eq.\
(\ref{OSDVlim}). The two-sided result $\Phi(t) \sim t^{-\gamma}$
is well known and  has been interpreted as an ``avoided crossing
effect'': trapping cannot be more efficient that the probability
$\int_t^\infty \psi(t) dt\sim t^{-\gamma}$ of remaining at the
initial site (see \cite{Klafter,BKZRev,Zumofen} and references
therein).  However, note that this ``avoided crossing effect'' is
absent for the one-sided trapping problem since the probability of
trapping $\Phi(t) \sim t^{-\gamma/2}$ decays even more slowly than
the probability of remaining at the initial site.

\subsection{Short-time behavior of $\Phi(t)$}
\begin{figure}
\includegraphics[width=0.95 \columnwidth]{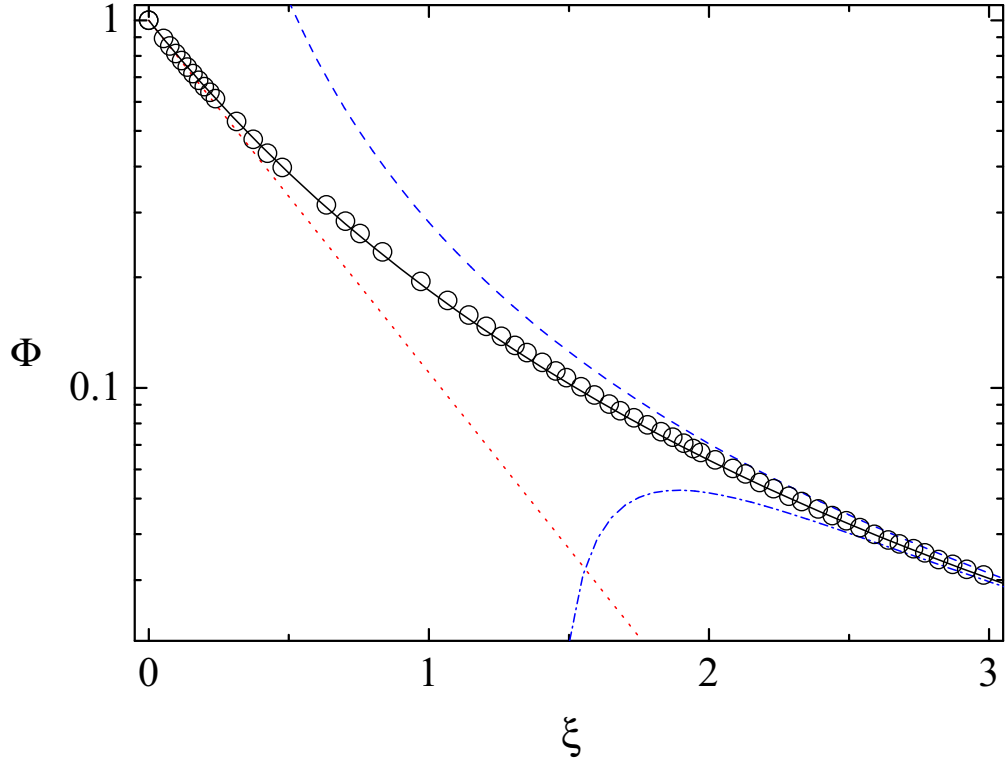}
\caption{Survival probability $\Phi(t)$ versus $\xi=\lambda
\sqrt{K_\gamma t^\gamma}$  for subdiffusive random walkers with
$\gamma=1/2$ for the two-sided trapping case. The symbols are
simulation results for $c=0.01$ and  $5\times 10^4$ trials. The
solid line corresponds to the numerical integration of the exact
expression \eqref{Phitavg}. The dashed and dash-dotted line are
the long-time behavior predictions obtained by keeping a single
term and three terms in Eq.\ (\protect\ref{Philarget}),
respectively. The dotted line is the short-time behavior given by
Eq.\ (\ref{Phitshort}).\label{fig2}}
\end{figure}

A simple analytical expression for the short-time behavior of
$\Phi(t)$ can also be  derived. To do so, we calculate the time
derivative of $\Phi_L(t)$ in Eq.\ (\ref{PhiL}) obtaining
\begin{equation}
\label{PhiLd}
\frac{d}{dt}\Phi_L(t)=- \frac{8}{L^2} K_\gamma
~_{0}D_{t}^{1-\gamma} J(\gamma)\; ,
\end{equation}
where we have taken into account that Mittag-Leffler functions are the solution of the fractional relaxation
equation \cite{MetzRev} and
\begin{equation}
\label{Jgam} J(\gamma)=\displaystyle\sum_{n=0}^\infty
E_\gamma\left(-4 a^2 (n+1/2)^2 \right)\;
\end{equation}
with $a^2=K_\gamma \pi^2 t^\gamma/L^2$. The major contribution to
the sum defining $J(\gamma)$ in Eq.\ (\ref{Jgam}) when $a$ is
small comes from large $n$. Hence, to find the lowest-order term
in the small-$a$ behavior of $J(\gamma)$, we can approximate the
sum by an integral:
\begin{equation}
\label{Jgamint} J(\gamma)\sim \int_0^\infty E_\gamma[-4 a^2 x^2]
dx = \frac{1}{2a} \widetilde E_\gamma(u=0)\; ,
\end{equation}
where $\widetilde E_\gamma(u)=\int_0^\infty e^{-u y} E_\gamma(y^2)
dy $ is the Laplace transform of $E_\gamma(y^2)$. To calculate
$\widetilde E_\gamma(u)$ we will exploit the relation between the
Mittag-Leffler and the Fox $H$ functions \cite{MetzRev,Mathai}:
\begin{equation}
\label{EgamH} E_\gamma(-y^2)=\frac{1}{2} H^{11}_{12}\left[y
\left|\begin{array}{l}{(0,1/2)}\\[1ex]{(0,1/2),(0,\gamma/2)}\end{array}\right. \right] \; .
\end{equation}
From Eq.\ (\ref{EgamH}) and the properties of Fox functions
\cite{West,Mathai} we find
\begin{equation}
\label{EgamHmore} \widetilde E_\gamma(s)=\frac{1}{2} H^{21}_{22}
\left[ s \left| \begin{array}{l}
{(1/2,1/2),(1-\gamma/2,\gamma/2)}\\[1ex] {(0,1),(1/2,1/2)}
\end{array} \right.\right] \; .
\end{equation}
Inserting the result in Eq.\ (\ref{EgamHmore}) into Eq.\
(\ref{Jgamint}), we finally arrive at the following explicit
expression for $J(\gamma)$:
\begin{equation}
\label{Jgamfin} J(\gamma)\sim\displaystyle\frac{1}{4
a}\displaystyle\frac{\pi}{\Gamma(1-\gamma/2)}=\displaystyle\frac{1}{4
\Gamma(1-\gamma/2)} \displaystyle\frac{L}{\sqrt{K_\gamma}}\,
t^{-\gamma/2} \; ,
\end{equation}
where we have calculated the value of the Laplace transform for $s=0$ by resorting to the general series
expansion of Fox functions \cite{MetzRev}. From Eqs.\ (\ref{PhiLd}) and (\ref{Jgamfin}) we can also write
\begin{equation}
\label{PhiLdfin}
\displaystyle\frac{d}{dt}\Phi_L(t)\sim-\displaystyle\frac{2}{\Gamma(\gamma/2)}
\displaystyle\frac{\sqrt{ K_\gamma}}{L} \, t^{\gamma/2-1}\; ,
\end{equation}
where the fractional derivative has been performed according to the rule: ${}_0D_t^\nu\, t^\mu=
\Gamma(1+\mu) \, t^{\mu-\nu}/ \Gamma(1+\mu-\nu)$ which is valid for arbitrary real parameters $\mu$
and $\nu$ \cite{MetzRev}. By integrating Eq.\ (\ref{PhiLdfin}) with the initial condition $\Phi_L(t=0)=1$
we find
\begin{equation}
\label{PhiLshort} \Phi_L(t)\sim1-
\frac{4}{\gamma\Gamma(\gamma/2)}\frac{\sqrt{K_\gamma
t^\gamma}}{L}\; ,
\end{equation}
and performing the average in Eq.\ (\ref{Phitavg}) over trap
configurations we finally obtain the short-time behavior of the
survival probability of a subdiffusive random walker in the
two-sided trapping problem
\begin{equation}
\label{Phitshort0} \Phi(t) \sim 1- \displaystyle\frac{4\xi}{\gamma
\Gamma(\gamma/2)}  \; .
\end{equation}
or, for $\xi \ll 1$,
\begin{equation}
\label{Phitshort} \Phi(t)  \sim \exp \left( -
\displaystyle\frac{2\xi}{\Gamma(1+\gamma/2)}  \right) \;.
\end{equation}
This expression is just the zeroth-order Rosenstock approximation
$\Phi(t)=\exp[-\lambda \langle S_\leftrightarrow(t) \rangle] $
[see Eq.\ (\ref{Ros:gen})] because the average two-sided span
$\langle S_\leftrightarrow(t) \rangle$ is simply twice the average
one-sided span $\langle S(t) \rangle$ and, by
Eq.~(\ref{SExploram}), $\langle S(t) \rangle=(K_\gamma
t^\gamma)^{1/2}/\Gamma(1+\gamma/2)$. In Fig.\ (\ref{fig2}) we
compare simulation results for subdiffusive random walkers in the
case $\gamma=1/2$ with the numerical integration of the exact
integral representation of $\Phi(t)$ given by Eq.\
(\ref{Phitavg}), the long-time behavior given in Eq.\
(\ref{Philarget}), and the Rosentock short-time behavior in Eq.\
(\ref{Phitshort}).

\section{CONCLUSIONS AND REMARKS}
\label{sec:conclusions} In this paper we considered the
one-dimensional trapping problem for subdiffusive particles. In
this problem a single subdiffusive random walker starting at $x=0$
on a lattice randomly filled with absorbing static traps with a
density $c$ performs a random exploration until he encounters one
of these traps. The quantity of main interest is the survival
probability $\phi(t)$ of the particle. We have considered two
versions of the problem: the one-sided case in which only one
half-line is filled with traps, and the two-sided case
corresponding to a random filling of both sides of the line $x >
0$ and $x < 0$. In the context of ordinary diffusive random
walkers, this problem has a long tradition, and many applications
to physics and chemistry have been discussed
\cite{Hughes,Weiss,HollanderWeiss,ShDba}.

Great interest has also arisen recently around subdiffusive
anomalous processes (for example, as a way of mimicking transport
in disordered media), and a continuous fractional diffusion
description has been put forward (see \cite{MetzRev} and
references therein).It thus seems convenient to extend the
fruitful trapping model to the case of subdiffusive particles. We
achieved this objective for the trapping of a single subdiffusive
random walker in one dimension and, by means of the fractional
diffusion formalism, exact analytical expressions were found in
terms of the special functions characteristic of fractional
calculus. There is a possibility for these results to be checked
in diffusion experiments performed in constrained geometries
and/or disordered materials \cite{Kosztolowicz}. From another
point of view, the trapping problem could also be interpreted as a
pseudo-first-order reaction of the form $A+T \rightarrow T$, where
$A$ is the random walker and $T$ is the trap. There is also an
increasing interest in reaction-subdiffusion processes
\cite{SKPRL,SKChemPhys}, and our exact results apply  to a special
class of these processes.

The long-time behavior of the survival probability $\Phi(t)$ has
been a subject of  particular interest in the ordinary diffusive
trapping problem \cite{Hughes,Weiss,DV,Anlauf}. In this regime an
anomalous fluctuation slowdown, known as the Donsker-Varadhan
limit, has been reported. In the one-sided and two-sided
subdiffusive models we  found the algebraic decays $\Phi(t) \sim
t^{-\gamma/2}$ and $\Phi(t) \sim t^{-\gamma}$, respectively. For
normal diffusive particles, the algebraic decay $\Phi(t) \sim
t^{-1/2}$ was found in \cite{Multipart} for the one-sided trapping
case and interpreted as due to the long explorations carried by
the random walkers on the half-line free of traps
\cite{Multipart}. A stretched exponential decay is the
corresponding behaviour for the long-time two-sided trapping of
diffusive particles \cite{Anlauf}. The two-sided subdiffusive
result $\Phi(t) \sim t^{-\gamma}$ is well known and  has been
interpreted as an ``avoided crossing effect'' (see
\cite{Klafter,BKZRev,Zumofen} and references therein). This effect
is absent for the one-sided trapping problem since the probability
of trapping $\Phi(t) \sim t^{-\gamma/2}$ for this case decays even
more slowly than the probability of remaining at the initial site.

The present work can continue along several directions. First, the
generalization of the subdiffusive trapping problem to the case of
$N > 1$ independent subdiffusive random walkers could be analyzed
as has already been done in the diffusive case \cite{Multipart}.
Application of the same techniques to higher dimensional spaces is
also an obvious approach, but we consider it unlikely to produce
exact closed expressions for the relevant quantities.  Trapping
models of subdiffusive particles in which the trapping process is
stochastically ``gated'' is also an interesting field. The effect
of ``gates'' in diffusion-limited reactions has recently been
considered in connection with reactions of complex molecules in
biological media \cite{PREBenichou}. The chemically active groups
of these molecules may be screened by the inactive parts, giving
rise to effective reactivities described by Poisson processes. The
media where these reactions takes place are usually disordered and
reaction-subdiffusion models should provide a better description
of these processes.

%\acknowledgments
This work was supported by the Ministerio de
Ciencia y Tecnolog\'{\i}a (Spain) through Grant No. BFM2001-0718
and by the European Community's Human Potential Programme under
contract HPRN-CT-2002-00307, DYGLAGEMEM.

%%%REFERENCES%%%%%%RRRRRRRRRRRRRRRRRRRRRRRRRRR%%%%%%%%%%%%%%

%FFFFFFFFFFFFFFFFFFFFFFFFFFFFFFFFFFFFFF

\end{document}